# A reduced transfer equation in one-dimensional spherical geometry with central symmetry

Charles H Aboughantous[*]

*Votan, 4548 Tigerland Avenue, Baton Rouge, Louisiana 70820, USA*

**Abstract.** The transfer equation in spherical geometry with central symmetry, which is a partial differential equation in two variables $r$ and $\mu$, is reduced to a one-dimensional transfer equation in $r$ parametric in $\mu$. This is done by dropping the angular derivative from the equation in conservation form without impeding the moments of the specific intensity. The justification for this reduction of the transfer equation is demonstrated analytically and numerically. The numerical demonstration is made by comparing the moments, zeroth to fourth, obtained using the analytic solutions in $r$ of both the reduced equation and the transfer equation in conservation form using the discrete ordinates method. The agreement between the two sets of data is perfect within round-off errors.

## 1. Introduction

Numerous astrophysical problems are addressed by solving the radiative transfer equation in spherical geometry with central symmetry. A number of numerical methods were proposed for that purpose [1]; some semi-analytical methods were also developed where the angular derivative is represented by approximations and the solution in $r$ is carried out analytically [2] [3]. The principal difference between the various solutions is the representation of the angular derivative. We devote this paper to a novel approach that dispense from the need of tackling with the angular derivative. Considering that the solution of the transfer equation is not aimed to obtain values for the specific intensity for its own sake, we propose to reduce the transfer equation by dropping out the angular derivative from the equation without impeding the moments. At first sight, this approach appears to be non-orthodox. We justify its validity by an analytical proof and by comparing numerical values of the moments computed from the analytical solutions of both the reduced transfer equation and the equation of transfer in conservation form with the angular derivative.

## 2. The reduced transfer equation

Consider the case of a homogeneous sphere of a gray medium in local thermodynamic equilibrium. The sphere is characterized by its radius $R$ and a source term defined as $\kappa B$, where $\kappa$ is the absorption coefficient and $B$ Plank's function. The standard form of the transfer equation is given as:

$$\mu \partial_r \psi + \frac{\eta^2}{r} \partial_\mu \psi + \kappa \psi = \kappa B \qquad (2.1)$$

where $\partial_r$ is the tensor notation of the derivative with respect to $r$, $\eta^2 = (1 - \mu^2)$ and $\psi \equiv \psi(r,\mu)$ is the specific intensity of monochromatic radiation. This equation is subject to the natural boundary conditions defined by a uniformly distributed intensity $\psi_R^-$ at the surface of the sphere and

---

[*] *Email address: votan@sprintmail.com*

symmetry at the center. Adding and subtracting $2\mu\psi/r$ to the left side of (2.1) yields:

$$\frac{\mu}{r^2}\partial_r(r^2\psi) + \frac{1}{r}\partial_\mu(\eta^2\psi) + \kappa\psi = \kappa B \tag{2.2}$$

Integration of (2.2) on $\mu \in [-1, 1]$ yields the first moment equation:

$$\frac{1}{r^2}\partial_r(r^2 F) + \kappa\varphi = \kappa B \tag{2.3}$$

where $\varphi \equiv \varphi(r)$ is the zeroth moment, or the total intensity, and $F \equiv F(r)$ is the first moment, or the flux. Higher moments equations can also be obtained. For instance, multiplying (2.1) by $\mu$ then adding and subtracting $(2\mu^2\psi/r + \eta^2\psi/r)$ to the left side yields:

$$\frac{\mu^2}{r^2}\partial_r(r^2\psi) - \frac{\eta^2}{r}\psi + \frac{1}{r}\partial_\mu(\mu\eta^2\psi) + \kappa\mu\psi = \kappa\mu B \tag{2.4}$$

Integration of (2.4) on $\mu \in [-1, 1]$ yields the second moment equation:

$$\partial_r K + \frac{3}{r}K + \kappa F - \frac{1}{r}\varphi = 0 \tag{2.5}$$

where $K \equiv K(r)$ is the second moment. One may continue this procedure to higher moments where the $m$th moment is defined as:

$$f_m = \frac{1}{2}\int_{-1}^{+1}\mu^m\psi\,d\mu \tag{2.6}$$

where $f_0 = \varphi$, $f_1 = F$, $f_2 = K$, etc.

It is apparent from (2.3) and (2.5) that any set of moments equations is coupled and it is open. Therefore, a rigorous solution of the moments equations may not be possible but a solution accurate to a preselected threshold can be obtained [4]. An alternative approach is to work with the first moment equation (2.3). Instead of solving for the moments directly, we solve for the *degenerate intensity* $\psi$, which is solution for the *angular degenerate* of (2.3):

$$\frac{\mu}{r^2}\partial_r(r^2\psi) + \kappa\psi + f(r,\mu) = \kappa B \tag{2.7}$$

where $f(r,\mu)$ is an odd function in $\mu \in [-1, +1]$ but otherwise arbitrary, and $\psi$ is some function of $r$ and $\mu$, not necessarily a specific intensity. In the particular case where $f(r,\mu) = r^{-1}\partial_\mu(\eta^2\psi)$, then (2.7) is identically the same as (2.2) and $\psi$ is the specific intensity of (2.2). It follows from (2.7) that the degeneracy of (2.3) on the angular domain is not unique, this is understood in light of the arbitrariness of $f(r,\mu)$.

*Theorem*: If $f(r,\mu)$ is any real valued odd function in $\mu \in [-1, +1]$ but otherwise arbitrary function, then the moments of $\psi$ solution of (2.7) are invariably the same moments of $\psi$ solution of (2.2) if and only if (2.7) is constrained to the same boundary conditions of (2.2),

*Proof*. Consider the case of the two equations (2.2) and (2.3). They both satisfy the same boundary conditions $\mathcal{B}$ and the moments of $\psi$ are $\{\varphi, F\}$, where $F$ is the flux and $\varphi$ the total intensity. Let (2.7) be subject to an arbitrary set of boundary conditions $\mathcal{B}'$ different from $\mathcal{B}$. Then, its angu-lar integral is an equation formally the same as (2.3) but now it satisfies the boundary conditions $\mathcal{B}'$. Therefore, the moments of $\psi$ are $F$ and $\varphi$ of (2.3) but they are not in general flux and intensity. Considering that $\mathcal{B}'$ is arbitrary, we may choose $\mathcal{B}' \equiv \mathcal{B}$. With this choice of boundary conditions, the angular integral of (2.7) is the same moment equation (2.3) that is obtained from (2.2), that is, now $F$ and $\varphi$ have the physical meanings of flux and intensity. Therefore, if $\{\varphi, F\}$ are the moments of $\psi$ solution of (2.2), they are also the moments of $\psi$ if and only if (2.7) is subject to the same boundary condition of (2.2).

Consider the second moment equation (2.5). If $F$ and $\varphi$ are moments of $\psi$ satisfying the boundary conditions $\mathcal{B}'$, then $K$ is also a moment of $\psi$ satisfying $\mathcal{B}'$. Therefore, if $\{\varphi, F, K\}$ are the moments of $\psi$, then $\{\varphi, F, K\}$ are invariably the same moments of $\psi$ if and only if $\psi$ satisfies the boundary condition $\mathcal{B}' \equiv \mathcal{B}$.

The proof of the theorem may be completed by recursion for all moments $\{f_m : m = 0, 1, \ldots\}$.

The validity of the theorem is exclusive to the moments of the specific intensity. It is obvious for that matter that $\psi \neq \psi \, \forall \mu$. Consequently if for any reasons the specific intensity is needed, or the half-range moments $f_m^\pm$ are needed for their own sake, (2.7) offers little help except for the particular case of $f(r,\mu) = r^{-1}\partial_\mu(\eta^2\psi)$, which means in effect we are working with (2.2).

Our motive to work with the degenerate transfer equation is to bypass working with the angular derivative which introduces numerical round-off of its own. This problem can be resolved by choosing a simple odd function $f(r,\mu)$. The simplest of all functions is $f(r,\mu) = 0$, a choice that leads to the simplest degenerate equation:

$$\frac{\mu}{r^2}\partial_r\left(r^2\psi\right) + \kappa\psi = \kappa B \qquad (2.8)$$

We dub (2.8) the *reduced transfer equation* and $\psi$ is the *reduced intensity*. It is noticeable that (2.8) can be obtained by simply dropping the angular derivative from (2.2).

In light of the foregoing, the advantage in working with the reduced transfer equation is compelling. The rest of this paper is devoted to the solution of (2.8) and pertinent analysis. Further formal simplification of (2.8) may be obtained by defining the function $\widetilde{\psi}_r = r^2\psi_r$. Then, the transformed reduced equation reads:

$$\partial_r\widetilde{\psi} + \lambda\widetilde{\psi} = \lambda r^2 B \qquad (2.9)$$

where $\lambda = \kappa/\mu$, that is, $\lambda$ is constant on the spatial domain; the case of a heterogeneous sphere will be examined in a later section.

We note that (2.9) is a partial differential equation. This is understood considering that $\widetilde{\psi}$ is function of $r$ and $\mu$. Therefore, a set of two boundary conditions is needed in order to obtain the complete solution for (2.9). The first boundary condition will be designated by $\widetilde{\psi}_R^- = R^2\psi_R^-$ where $\psi_R^-$ is *natural boundary condition*. The second boundary condition is problem dependent, as illustrated in later sections, but it always boils down to a required symmetry at the center of the sphere consistently with the definition of the problem addressed in this paper.

## 3. The two-streams reduced equations

The solution of (2.9) is straightforward and the natural boundary condition can be applied directly at the surface of the sphere for $\mu < 0$. The application of the second boundary condition requires splitting the solution into two-streams solution, one for $\mu < 0$ and one for $\mu > 0$. Instead, we find it convenient to use Schuster's two-streams technique [5]. Let $\psi^+$ be a centrifugal intensity and $\psi^-$ a centripetal intensity in a direction defined by $\mu \in [0, 1]$. Then the two-streams equations read:

*Centripetal equation:* $\qquad \partial_r \widetilde{\psi}^- = \lambda \widetilde{\psi}^- - \lambda r^2 B$ (3.1)

*Centrifugal equation:* $\qquad \partial_r \widetilde{\psi}^+ = -\lambda \widetilde{\psi}^+ + \lambda r^2 B$ (3.2)

Considering that $r > 0$, the spatial domain of definition of these two equations is $[\varepsilon > 0, R]$ where $\varepsilon$ is a radius of a concentric spherical surface. The $\varepsilon$-surface defines a *pellet* that is thermodynamically different from the sphere and it will be characterized by its absorption coefficient $\kappa_p$ and its Plank function $B_p$; the case of $\varepsilon = 0$ will be discussed in due course. Consequently, the two equations (3.1) and (3.2) can be solved with the mathematical boundary conditions $\psi_\varepsilon^-$ and $\psi_\varepsilon^+$, yet to be determined in terms of the natural boundary condition $\psi_R^-$.

We observe that (3.1) and (3.2) are symmetric. Therefore, one should expect that their solutions generate the same moments of the symmetrized solutions of the nonsymmetrized discrete ordinates equations of [3]. The first confirmation to such an expectation is that both, (3.1) and (3.2) have one common solution with the symmetrized solutions of the discrete ordinates equations. In vacuum, we would have: $\psi_r^\pm = \psi_r^\mp = (R/r)^2 \psi_R^-$, which expresses the reciprocity of path.

Once the solutions $\psi^\pm$ are obtained, their moments are computed using (2.6). That integral cannot be carried out in closed form. The recourse is a numerical integration. We opt for Gauss-Legendre quadrature:

$$f_m = \frac{1}{2} \sum_{n=1}^{N} w_n \mu_n^m \left( \psi_n^+ + (-1)^m \psi_n^- \right)$$ (3.3)

The choice of this quadrature is particularly instructive because it is also used in the discrete ordinates method [3]. Therefore, the round-off error from the quadrature does not make a part of discrepancies between the moments computed from the reduced transfer equation and those computed using the discrete ordinates method.

### 3.1. Solution of the centripetal equation

The solution of (3.1) on the domain $[\varepsilon, R]$ can be obtained directly in terms of the mathematical boundary condition. Evaluate the solution at $R$ and rewrite it to read:

$$\psi_\varepsilon^- = \frac{R^2}{\varepsilon^2} e^{-\lambda(R-\varepsilon)} \psi_R^- + \frac{B}{\varepsilon^2} \left[ Q_\varepsilon^- - Q_R^- e^{-\lambda(R-\varepsilon)} \right]$$ (3.4)

$$Q_x^- = x^2 + \frac{2x}{\lambda} + \frac{2}{\lambda^2}$$ (3.5)

where $\psi_R^-$ is the natural boundary condition. It is apparent from (3.4) that $\varepsilon$ can be treated as an independent parameter and eventually an independent variable $r$. The complete expression for the centripetal intensity for $r \in [\varepsilon, R]$ becomes:

$$\psi_r^- = \frac{R^2}{r^2} e^{-\lambda(R-r)} \psi_R^- + \frac{B}{r^2}\left[Q_r^- - Q_R^- e^{-\lambda(R-r)}\right] \tag{3.6}$$

### 3.2. Solutions of the centrifugal equation

Direct integration of (3.2) using the mathematical boundary condition yields:

$$\psi_r^+ = \frac{\varepsilon^2}{r^2} e^{-\lambda(r-\varepsilon)} \psi_\varepsilon^+ + \frac{B}{r^2}\left[Q_r^+ - Q_\varepsilon^+ e^{-\lambda(r-\varepsilon)}\right] \tag{3.7}$$

$$Q_x^+ = x^2 - \frac{2x}{\lambda} + \frac{2}{\lambda^2} \tag{3.8}$$

The search for the boundary value $\psi_\varepsilon^+$ in terms of the natural boundary condition $\psi_R^-$ depends on the optical processes that take place at $\varepsilon$-surface. We consider a few examples.

*(a) Specular reflection.* In this case, the reflection is perfect at $\varepsilon$-surface and it obeys Snell's law: $\psi_\varepsilon^+ = \psi_\varepsilon^-$ where $\psi_\varepsilon^-$ is obtained from (3.4). Then (3.7) reads:

$$\psi_r^+ = \frac{R^2}{r^2} e^{-\lambda(R+r-2\varepsilon)} \psi_R^- + \frac{B}{r^2}\left[\frac{4\varepsilon}{\lambda} e^{-\lambda(r-\varepsilon)} - Q_R^- e^{-\lambda(R+r-2\varepsilon)} + Q_r^+\right] \tag{3.9}$$

This solution is valid if $\varepsilon$-surface is a convex spherical mirror, or the pellet is merely a cavity. In the latter case, the reflection is assumed at the center and $\psi_{r\leq\varepsilon}^+ = \psi_{r\leq\varepsilon}^-$ because of the principle of reciprocity of path within the cavity.

*(b) Albedo boundary condition.* Evaluate (3.9) at $R$:

$$\psi_R^+ = e^{-2\lambda(R-\varepsilon)} \psi_R^- + \frac{B}{R^2}\left[\frac{4\varepsilon}{\lambda} e^{-\lambda(R-\varepsilon)} - Q_R^- e^{-2\lambda(R-\varepsilon)} + Q_R^+\right] \tag{3.10}$$

It is apparent from this expression that the operation of the coefficient of $\psi_R^-$ on its operand bears the significance of *albedo* of a spherical shell of thickness $(R - \varepsilon)$. This albedo has little in common with the traditional definition of albedo as the ratio of the entering to the exiting radiation at a surface driven by scattering within the sphere. We do not have scattering in the present case. That is, the albedo $A_{R,\varepsilon} = e^{-2\lambda(R-\varepsilon)}$ is merely a manifestation of symmetry.

We obtain the albedo of a solid sphere of radius $R$ by setting $\varepsilon = 0$ in (3.10): $A_R = e^{-2\lambda R}$. Hence, (3.10) can be resolved to express the centrifugal intensity at the surface of the pellet:

$$\widehat{\psi}_\varepsilon^+ = e^{-2\lambda_p \varepsilon} \psi_\varepsilon^- + \frac{B_p}{\varepsilon^2}\left[Q_\varepsilon^+ - Q_\varepsilon^- e^{-2\lambda\varepsilon}\right]_p \tag{3.11}$$

where the quantities with subscript $p$ pertain to the pellet and $\psi_\varepsilon^-$ is given by (3.4). The intensity $\widehat{\psi}_\varepsilon^+$ of (3.11) can now be used for the boundary value $\psi_\varepsilon^+$ of (3.7) to obtain the solution for the centrifugal equation with albedo boundary condition.

At the limit as $\varepsilon \to 0$ we should expect $\widehat{\psi}_{\varepsilon\to 0}^+ \to \psi_0^+ = \psi_0^-$, but (3.11) yields:

$$\psi_0^+ = \psi_0^- - \left.\frac{4B_p}{\lambda_p \varepsilon}\right|_{\varepsilon \to 0} \tag{3.12}$$

This is manifestly contrary to the expectations: symmetry is expected at the center but (3.12) suggests otherwise. The missing information from (3.12) is that the limit term is undetermined: as $\varepsilon \to 0$ the pellet vanishes and drags $B_p$ and $\kappa_p$ with it to naught; one has to keep in mind that both $B_p$ and $\kappa_p$ are meaningful only in a statistical sense in the interior of a non-vanishing material volume. The apparent inconsistency of (3.12) illustrates the fact that the linearized transfer equation (2.1) is not valid at and in the close proximity of the center.

*(c) Mixed boundary condition.* A great number of physical situations may be described by this type of boundary condition, particularly in radiative transfer of planetary atmospheres [6]. We illustrate the case of a partially reflective $\varepsilon$-surface. Let $\alpha$ be the fraction of the incident radiation that is reflected at the surface, then the reflected intensity is $\alpha \psi_\varepsilon^-$ and the albedo intensity at the surface of the pellet becomes:

$$\widehat{\psi}_\varepsilon^{\alpha+} = (1-\alpha) e^{-2\lambda_p \varepsilon} \psi_\varepsilon^- + \frac{B_p}{\varepsilon^2}\left[Q_\varepsilon^+ - Q_\varepsilon^- e^{-2\lambda \varepsilon}\right]_p \tag{3.13}$$

The mixed boundary intensity becomes:

$$\ddot{\psi}_\varepsilon^+ = \widehat{\psi}_\varepsilon^{\alpha+} + \alpha \psi_\varepsilon^- \tag{3.14}$$

where $\psi_\varepsilon^-$ of (3.13) and (3.14) is obtained from (3.4) and the mixed boundary intensity $\ddot{\psi}_\varepsilon^+$ of (3.14) takes the place for $\psi_\varepsilon^+$ of (3.7) to complete the solution for the centrifugal intensity with this type of boundary condition.

### 3.3. Analytical properties of the solutions

First, we look at the intensity in the asymptotic region. Whichever boundary condition is applied at the surface of the pellet we obtain the diffusion regime: $r = R \to \infty \Rightarrow \psi^+ = \psi^- = \psi^+ = \psi^- = B$.

Define $\phi = (\psi^+ + \psi^-)/2$ and its real conjugate $\phi^* = (\psi^+ - \psi^-)/2$, these are commonly known as Feautrier parameters [7]. Historically, Schuster [5] seems to be the first to define $\phi$ as the mean intensity. Kozirev [8] defined $\phi$ as the total intensity in the direction corresponding to the root-mean-square of direction cosine $\mu = 3^{-1/2}$ and credited this to Eddington. Vladimirov [9] defined $\phi$ and $\phi^*$ as even and odd parity functions in $4\pi$-steradians in his treaties on theory of particle transport and he credited them to Kuznetzov [10].

It is obvious from the expressions for the reduced intensities that the total intensity is singular at the center of the sphere. The property of the flux may be examined by analyzing the property of $\phi^*$ in the case of reflective boundary condition, and $\phi^{*\alpha}$ in the case of albedo boundary condition. Without loss of generality, we consider the case of a solid homogeneous sphere with a concentric surface of radius $\varepsilon$ so that the interior of the pellet and the medium of the sphere have the same properties $B$ and $\kappa$. We seek reflective and albedo boundary conditions at $\varepsilon$. Hence:

$$\phi_r^* = -\frac{R^2}{r^2}\psi_R^- e^{-\lambda(R-\varepsilon)}\sinh\lambda(r-\varepsilon) + \frac{B}{r^2}\left[\frac{2\varepsilon}{\lambda}e^{-\lambda(r-\varepsilon)} - \frac{2r}{\lambda} + Q_R^- e^{-\lambda(R-\varepsilon)}\sinh\lambda(r-\varepsilon)\right] \tag{3.15}$$

$$\phi_r^{*\alpha} = -\frac{R^2}{r^2}\psi_R^- e^{-\lambda R}\sinh\lambda r + \frac{B}{r^2}\left[-\frac{2r}{\lambda} + Q_R^- e^{-\lambda R}\sinh\lambda r\right] \qquad (3.16)$$

It is apparent from (3.15) that the flux vanishes at $r = \varepsilon$. However, the flux of the discrete ordinates method spikes before it vanishes at $r = \varepsilon$. The justification of the spike is demonstrated graphically in the case of a sphere without emission [3]. The expressions of the discrete ordinates solution are overly complex to carry out an analytic demonstration of the existence and the location of the spike. In contrast, the expression (3.15) is simple and manageable to determine the position $r_{sp}$ of the spike. Set $\partial_r \phi^* = B = 0$ and rearrange to obtain the $z$-function:

$$z(r) = \int_0^1 \left[\mu r - (2\mu/\lambda)\tanh\lambda(r-\varepsilon)\right] d\mu \qquad (3.17)$$

The zero of this $z$-function corresponds to the location of the spike. We note that an equivalent expression to (3.17) can be obtained with $B > 0$. However, for the purpose of explaining the spike behavior of the flux, (3.17) suffices. Now, set $\partial_r z(r) = 0$ and rearrange to obtain:

$$\int_0^1 \tanh^2(\ell/\mu)\, d\mu = \frac{1}{4} \qquad (3.18)$$

where the optical length $\ell = \kappa(r - \varepsilon)$. Numerical integration of (3.18) indicates that there exist only one value $\ell = 0.5493069$ for which the flux in a non-emissive sphere has a point of inflection. Therefore, the flux can have two extrema, or it is monotonic with one point of inflection, or monotonic with no point of inflection; there is a point of inflection if $\kappa \geq 0.55/(R - \varepsilon)$. Numerical integration of (3.17) should yield the locations of the extrema points for a given sphere. Clearly, this result is valid for a uniform $\kappa$ throughout a non-emissive sphere.

If we set $\varepsilon = 0$ in (3.15), we recover (3.16), which indicates that the flux is singular at the center of the sphere, except in vacuum. This is paradoxical since we must have $\psi_0^+ - \psi_0^- = 0$, which implies $F_0 = 0$ by reason of symmetry. The paradox is understood considering that $\psi_0^+$ and $\psi_0^-$ are undetermined and hence $\psi_0^+ = \psi_0^-$ cannot be justified. This result is consistent with the invalidity of the linearized transfer equation at and in the close proximity of the center. However, we note that both (3.15) and (3.16) yield $r^2 \phi_r^*\big|_{r=\varepsilon} = 0$ as $\varepsilon \to 0$. That is, although the flux is manifestly singular at the center, no flow of energy occurs through that point, which implies total reflection at the center of the sphere.

Another observation worth noting is the negative $2r/\lambda$ term inside the square brackets of (3.15) and (3.16). It suggests that the flux in the interior of emissive spheres immersed in vacuum could vanish at a point other than the center in the interior of the sphere. This null flux point defines an adiabatic surface its radius is independent of $B$ but it depends on $R$, $\varepsilon$ and $\kappa$.

## 4. The micropellet problem

The transfer equation is used as a diagnostic tool in inertial fusion of imploded micropellets. The energy density computed from the specific intensity, which is proportional to the zeroth moment of $\psi$, is then used as a source term for the hydrodynamics calculations of the imploded pellet. Considering that the pellet has a radius of 25 to 50 µm, if the specific intensity is sought too close to the center, assuming the validity of the linearized transfer equation at that proximity to the center, the terms $QB/r^2$ of the solutions are bound to return an overflow message. One way to bypass this problem is to define a spherical shell of outer radius equal to the diameter of the pellet and inner radius equal to the radius of the pellet.

The exact specific intensity in the shell can be obtained from the solutions discussed in the previous section for $r \in [R, 2R]$ by making the substitutions $\varepsilon \leftarrow R$ and $R \leftarrow 2R$ in the expressions for the solutions of the reduced equation. The boundary condition at the outer surface of the shell is the same $\psi_R^-$ of the micropellet. The inner surface of the shell coincides with the center of the micropellet, which is a reflective boundary. Once the specific intensity $\psi_s$ is computed in the shell, the total energy $E_s$ stored in the volume of the shell is computed from the integral of $\kappa_s \varphi_s$ on the volume of the shell. Then the energy in the micropellet is computed from $E_p = \alpha E_s$ where $\alpha$ is the ratio of the volume of the pellet to the volume of the shell. The same procedure applies to a stratum of the micropellet relative to a stratum of the shell. Detailed discussion on this subject can be found in [11].

## 5. The heterogeneous sphere

We considered the case of $\lambda$ constant on the spatial domain. If $\lambda(r)$ is completely prescribed and if it is integrable, then the integration of (3.1) and (3.2) is straightforward using standard methods. Most often, the medium is stratified into homogeneous strata. In that case, the solutions of the previous section apply to each stratum individually.

Consider a sphere stratified into $M$ strata not including the pellet and each stratum is designated by its index $m = 1, 2, ..., M$ so that the stratum adjacent to the pellet is designated by $m = 1$ and the outermost stratum by $M$. In this arrangement, the $m$th stratum is characterized by $\lambda_m$ and $B_m$ and it is bounded by $r_m > r_{m-1}$. Consequently, the boundary conditions for each stratum are $\psi_m$ and $\psi_{m-1}$. In order to determine the boundary conditions we use the solutions of the previous section with the following substitutions: $r_m \leftarrow R$ and $r_{m-1} \leftarrow \varepsilon$, $\psi_m^- \leftarrow \psi_R^-$ and $\psi_0^\pm \leftarrow \psi_\varepsilon^\pm$. Then, using the ART properties (Albedo, Radiance, Transmittance) [12], the solutions on the continuous variable $r$ transcribe into the end-points solutions:

$$\psi_{m-1}^- = T_{m-1}^m \psi_m^- + P_{m-1}^m \tag{5.1}$$

$$\psi_m^+ = T_m^{m-1} \psi_{m-1}^+ + P_m^{m-1} \tag{5.2}$$

where $T_x^y$ is the transmittance and $P_x^y$ the radiance of a stratum; the transport is from $y$ to $x$. The solutions of (5.1) start with $m = M$ for each direction defined by $\mu$ and continue with decreasing $m$. At $m = 1$, set $\psi_0^+ = \psi_0^-$ and proceed sweeping (5.2) with increasing $m$.

## 6. Quantitative analysis

We generated graphs for the moments of the specific intensities using the solution of the discrete ordinates equations [3] and the solution of the reduced transfer equation of this paper. The solution $\psi$ using the discrete ordinates method contains the contribution of the angular derivative to the specific intensity, the solution $\psi$ of the reduced equation does not. The computed specific intensities from the two solutions do not agree, as expected. In contrast, the moments of $\psi$ and the moments of $\psi$ are in excellent agreement. The graphical comparison of the moments is perfect, pixel for pixel. For the sake of avoiding redundancy, we do not reproduce the graphs here. The reader is invited to examine the graphs in ref. 3.

**Table 1.** Comparison of the moments, zeroth to fourth, computed using the discrete ordinates method and the reduced transfer method; the angular derivative takes part of the solution using the former method. The differences are absolute values at different radial positions in spheres of radius $R = 1.0$ hosting a reflective pellet of radius $\varepsilon = 0.1$. $\kappa$ is the opacity and $N$ is the order of the quadrature/discrete ordinates. The external radiation field is uniform and isotropic of normalized intensity. The sphere is assumed a gray medium without emission.

| | $\kappa = 0.1$ | | | $\kappa = 5.0$ | | |
|---|---|---|---|---|---|---|
| N | $r = 0.1$ | $r = 0.55$ | $r = 1.0$ | $r = 0.1$ | $r = 0.55$ | $r = 1.0$ |
| | | | *Total intensities*: $\Delta\varphi$ | | | |
| 1  | 0.0                 | $8.9\times10^{-16}$ | $1.1\times10^{-16}$ | $1.4\times10^{-17}$ | $2.1\times10^{-17}$ | $4.4\times10^{-16}$ |
| 2  | $2.8\times10^{-14}$ | $8.9\times10^{-16}$ | $2.2\times10^{-16}$ | 0.0                 | $2.1\times10^{-17}$ | $4.4\times10^{-16}$ |
| 4  | $1.4\times10^{-14}$ | $8.9\times10^{-16}$ | $1.1\times10^{-16}$ | $1.1\times10^{-16}$ | $2.1\times10^{-17}$ | $5.6\times10^{-16}$ |
| 8  | $1.4\times10^{-14}$ | $1.3\times10^{-15}$ | $1.1\times10^{-16}$ | 0.0                 | $2.1\times10^{-17}$ | $7.8\times10^{-16}$ |
| 16 | $1.4\times10^{-14}$ | $8.9\times10^{-16}$ | $1.1\times10^{-16}$ | $1.1\times10^{-16}$ | $1.4\times10^{-17}$ | $1.1\times10^{-15}$ |
| | | | *Fluxes*: $\Delta F$ | | | |
| 1  | $1.5\times10^{-15}$ | $1.1\times10^{-16}$ | $5.6\times10^{-17}$ | $1.5\times10^{-19}$ | $1.0\times10^{-17}$ | $1.7\times10^{-16}$ |
| 2  | $4.2\times10^{-16}$ | 0.0                 | $4.2\times10^{-17}$ | $2.7\times10^{-18}$ | $2.1\times10^{-17}$ | $1.7\times10^{-16}$ |
| 4  | $1.7\times10^{-15}$ | 0.0                 | $4.2\times10^{-17}$ | $2.7\times10^{-18}$ | 0.0                 | $3.3\times10^{-16}$ |
| 8  | $9.2\times10^{-16}$ | $9.7\times10^{-17}$ | $5.6\times10^{-17}$ | $2.8\times10^{-18}$ | $6.9\times10^{-18}$ | $2.8\times10^{-16}$ |
| 16 | $1.1\times10^{-15}$ | 0.0                 | 0.0                 | $5.9\times10^{-21}$ | $1.4\times10^{-17}$ | $1.7\times10^{-16}$ |
| | | | *Second moments*: $\Delta K$ | | | |
| 1  | $6.5\times10^{-13}$ | $2.0\times10^{-14}$ | $5.3\times10^{-15}$ | $2.7\times10^{-16}$ | $1.2\times10^{-16}$ | 0.0 |
| 2  | $3.5\times10^{-13}$ | $1.0\times10^{-14}$ | $2.8\times10^{-15}$ | $1.2\times10^{-15}$ | $2.1\times10^{-16}$ | 0.0 |
| 4  | $2.3\times10^{-13}$ | $7.3\times10^{-15}$ | $2.0\times10^{-15}$ | $7.8\times10^{-16}$ | $1.2\times10^{-16}$ | 0.0 |
| 8  | $5.0\times10^{-14}$ | $1.6\times10^{-15}$ | $5.0\times10^{-16}$ | $8.0\times10^{-16}$ | $1.0\times10^{-16}$ | 0.0 |
| 16 | $7.8\times10^{-14}$ | $2.7\times10^{-15}$ | $6.1\times10^{-16}$ | $8.3\times10^{-17}$ | $2.8\times10^{-17}$ | 0.0 |
| | | | *Third moments*: $\Delta L$ | | | |
| 1  | $1.2\times10^{-15}$ | $5.1\times10^{-15}$ | $3.1\times10^{-15}$ | $1.7\times10^{-19}$ | $6.4\times10^{-17}$ | 0.0 |
| 2  | $1.5\times10^{-17}$ | $3.5\times10^{-15}$ | $2.1\times10^{-15}$ | $2.8\times10^{-18}$ | $1.8\times10^{-16}$ | 0.0 |
| 4  | $1.3\times10^{-15}$ | $3.8\times10^{-15}$ | $2.2\times10^{-15}$ | $1.8\times10^{-18}$ | $9.7\times10^{-17}$ | $9.7\times10^{-16}$ |
| 8  | $9.4\times10^{-16}$ | $1.2\times10^{-15}$ | $5.5\times10^{-16}$ | $1.8\times10^{-19}$ | $9.0\times10^{-17}$ | 0.0 |
| 16 | $1.1\times10^{-16}$ | $8.3\times10^{-16}$ | $6.1\times10^{-16}$ | $4.8\times10^{-18}$ | $1.7\times10^{-17}$ | 0.0 |
| | | | *Fourth moment*: $\Delta M$ | | | |
| 1  | $2.1\times10^{-13}$ | $6.6\times10^{-15}$ | $1.8\times10^{-15}$ | $9.0\times10^{-17}$ | $3.7\times10^{-17}$ | 0.0 |
| 2  | $2.2\times10^{-13}$ | $6.6\times10^{-15}$ | $1.7\times10^{-15}$ | $9.0\times10^{-16}$ | $1.6\times10^{-16}$ | 0.0 |
| 4  | $2.5\times10^{-13}$ | $8.1\times10^{-15}$ | $2.3\times10^{-15}$ | $5.0\times10^{-16}$ | $8.0\times10^{-17}$ | 0.0 |
| 8  | $5.0\times10^{-14}$ | $1.7\times10^{-17}$ | $6.1\times10^{-16}$ | $6.0\times10^{-16}$ | $6.9\times10^{-17}$ | 0.0 |
| 16 | $3.9\times10^{-14}$ | $1.9\times10^{-15}$ | $5.0\times10^{-16}$ | $9.7\times10^{-17}$ | $1.7\times10^{-17}$ | 0.0 |

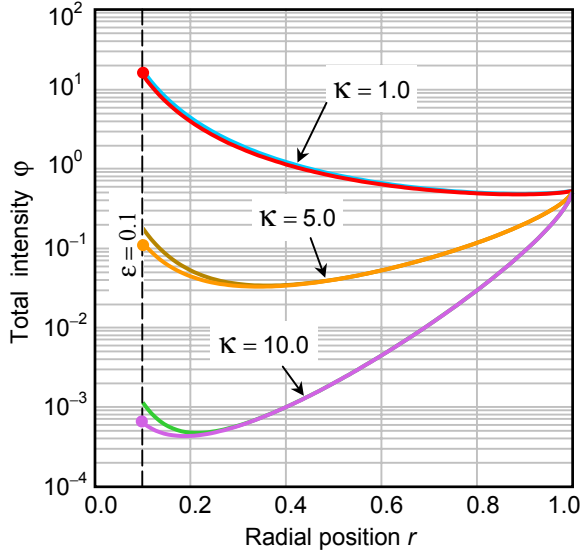

**Figure 6.1.** Graphs of intensities in solid homogeneous spheres of different opacities; $B = 0$. The graphs marked with a ball are generated with albedo boundary condition at $\varepsilon = 0.1$, the other graphs are generated with specular reflection boundary condition.

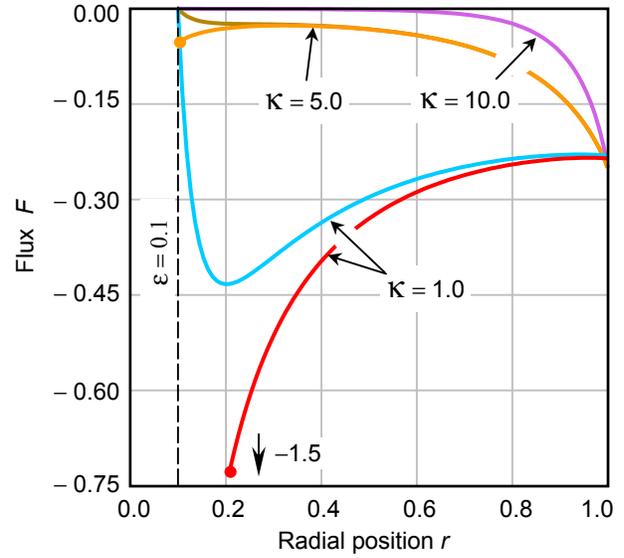

**Figure 6.2.** Graphs of fluxes in solid homogeneous spheres of different opacities; $B=0$. The graphs marked with a ball are generated with albedo boundary condition at $\varepsilon = 0.1$, the other graphs are generated with specular reflection boundary condition.

Numerical data for the moments, zeroth to fourth, using the discrete ordinates solution and the solution of the reduced equation were generated in two homogeneous spheres characterized by $B = 0$, $\kappa = 0.1$ and $\kappa = 5.0$. All computations were done with 16 significant digits arithmetics using equation (5.1) and (5.2) and their counterparts of the discrete ordinates method [3] with reflective boundary condition at the surface of the pellet. We compared the data at the surface of the pellet, at mid-point in the shell bounded by $\varepsilon = 0.1$ and $R = 1.0$, and at the surface of the sphere.

The differences between the reduced transfer moments are computed from $\psi$ and are designated with subscript *red*, the moments computed from $\psi$, the solution using the discrete ordinates method, are designated with subscript *disc*. The data are shown in Table 1 as $\Delta\varphi = |\varphi_{disc} - \varphi_{red}|$ for the zeroth moment, $\Delta F$ for the first moment, $\Delta K$ the second moment, $\Delta L$ third moment and $\Delta M$ the forth moment. Ideally, the differences as predicted by the theorem sec. 2 must be zero. The tabulated data show non-zero values but small enough to be treated as zeros within round-off error in the finite arithmetics of the computations. These results confirm the validity of the theorem. The CPU time ratio of the discrete ordinates computations to the reduced transfer computations increases with the order of discrete ordinates. The largest ratio measured with the data of Table 1 was about 140 to 1.

We generated three pairs of graphs for the total intensities and three pairs for the fluxes using the reduced transfer model in three solid homogeneous spheres of different opacities and without emission. A surface of radius $\varepsilon = 0.1$ is selected to simulate a pellet for the purpose of applying reflective and albedo boundary conditions at that location. Figure 6.1 shows the graphs of the total intensities and figure 6.2 shows the fluxes. The graphs that were generated with albedo boundary conditions terminate with a ball at $r = \varepsilon$, the other graphs were generated with reflective boundary condition. In all cases $\psi_R^- = 1.0$.

In the first sphere, $\kappa = 1.0$, the graphs for the intensity are nearly coinciding. This is expected since the radius of the pellet is only 0.1 mfp, the pellet behaves almost like a cavity: the reflected intensity is nearly the same as the transmitted intensity through pellet. The discrepancy between the graphs of each pair near the surface of the pellet increases with increasing $\kappa$. This is expected, all the energy that reaches the surface of the pellet is reflected at the surface, but only a fraction of it emerges from the pellet with albedo boundary condition, the other fraction is lost by absorption within the pellet. It is noticeable that the graphs in each pair coincide a short distance from the surface of the pellet. This is because most of the contribution to the total intensity in the region away from the surface of the pellet comes from the centripetal intensity; the centrifugal intensity faints out quickly relative to the centripetal intensity.

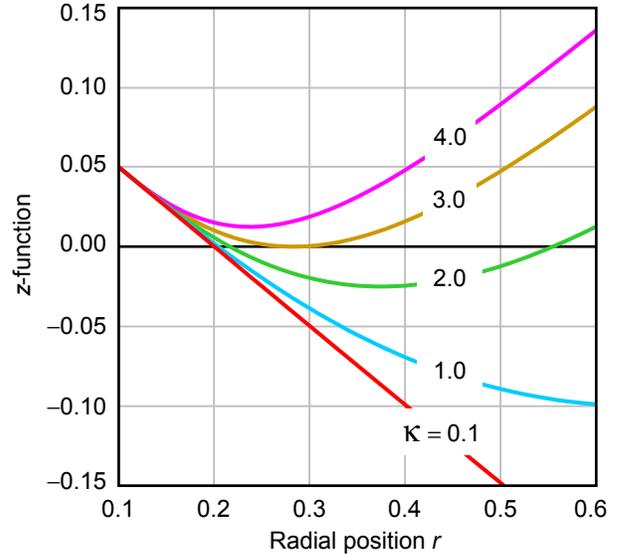

**Figure 6.3.** Variations of the z-function in spheres of different opacities $\kappa$. The radius of the spheres is $R = 1.0$ and the radius of the pellet $\varepsilon = 0.1$.

The graphs of the fluxes shown in figure 6.2 reveal the expected behavior: the fluxes computed with reflective boundary condition vanish at the surface of the pellet, the fluxes computed with albedo boundary condition do not. All fluxes are negative because there is no blackbody emission in these spheres. We experimented with $B_p > 0$ while maintaining $B_{sphere} = 0$. In those cases, the flux at the surface of the pellet becomes positive, then within a short distance in mfp units from $\varepsilon$-surface it becomes negative. That is, the incident flux is larger than the exiting flux at the surface of the sphere.

In all the cases of figures 6.1 and 6.2, if we change $\varepsilon$ the same graph is obtained with albedo boundary condition but only truncated at $\varepsilon$, it is a new graph with reflective boundary condition. In the latter case, changing $\varepsilon$ is indeed changing the thickness of the shell bounded by $\varepsilon$ and $R$, which is a different sphere. In the former case, changing $\varepsilon$ is merely moving $\varepsilon$-surface around in the interior of the same solid sphere.

We note that the graph corresponding to $\kappa = 1.0$ of figure 6.2 spikes very closely at $r_c = 0.2$ which is a zero of the z-function of (3.17). There is only one spike corresponding to this opacity as predicted from the solution of (3.18). The variations of the z-function are shown in figure 6.3 for different opacities. The graphs show the existence and the locations of the zeros in a sphere of normalized radius $R = 1$; the location of the point of inflection is determined at the minimum of the graphs.

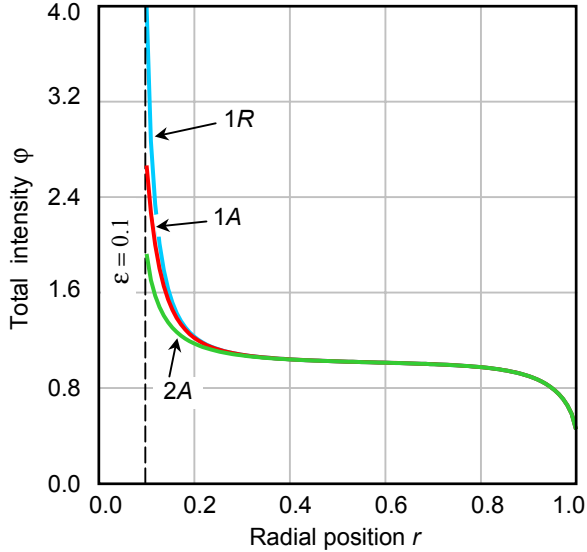

**Figure 6.4.** Graphs of total intensities in two spheres of different but uniform $B$ and variable opacities. Different boundary conditions, $R$ for reflection, $A$ for albedo, are applied at the surface of the pellet and vacuum boundary condition at the surface of the spheres.

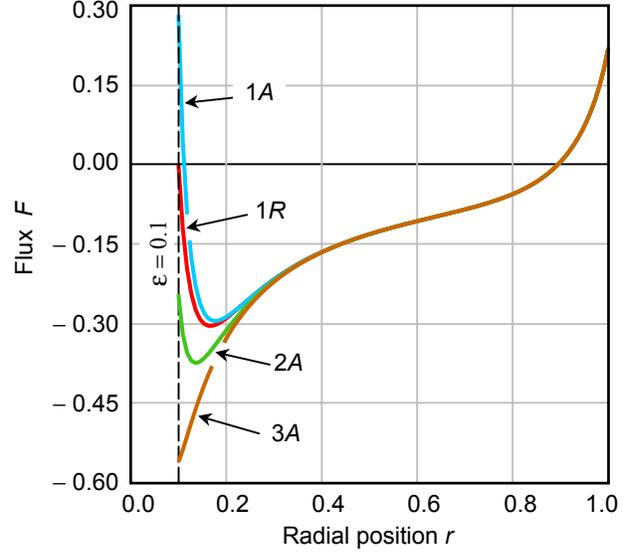

**Figure 6.5.** Graphs of fluxes in three spheres of different but uniform $B$ and variable opacities. Different boundary conditions, $R$ for reflection, $A$ for albedo, are applied at the surface of the pellet and vacuum boundary condition at the surface of the spheres.

We experimented with heterogeneous spheres made of homogeneous strata. The absorption coefficient of the pellet is fixed at $\kappa = 10$ and $\kappa_m = 0.1/r_m^2$ in the $m$th stratum in the sphere. Total intensities and fluxes were calculated in three spheres characterized with different blackbody constants: $B = 10$ in the first sphere, $B = 5$ in the second sphere and $B = 2$ in the third sphere; the surface boundary condition is $\psi_R^- = 0$ in all cases. The graphs shown in figure 6.4 and in figure 6.5 are labeled with a number, which is the ordinal of the sphere, and a letter: $A$ designates albedo boundary condition and $R$ designates a reflective boundary condition at the surface of the pellet. The graphs of figure 6.5 show that the spike of the flux persists for some values of $\kappa$, $\varepsilon$, $R$, $B$ of the sphere and $B$ of the pellet.

It is noticeable that the flux in figure 6.5 vanishes at the adiabatic surface near the surface of the sphere, as predicted in Sec. 3. In some cases, the flux vanishes at another surface close to the surface of the pellet, graph 1$A$, or at the surface of the pellet with reflective boundary condition, graph 1$R$. The instructive information from this pattern of the flux is that an observer outside the sphere would not obtain radiative information from the region of the sphere interior to the adiabatic surface. The energy that radiates out of the sphere appears to be originating from a shell no deeper than the adiabatic surface.

It was shown that the medium hosting a radiation field sustains a mechanical radiative pressure proportional to the flux [1]. Consequently, the graphs of the flux of figure 6.5 are represen-tative of the radiative forces on the sphere. It is apparent that the radiative energy tends to sputter the outer shell of the sphere while the inner shell is compressed with a maximum pressure at the spike location.

We experimented with stratified spheres characterized by inverse square opacities and inverse square emission law. The pattern of the graphs of figures 6.4 and 6.5 are representative of the total intensities and fluxes in all of these cases.

## 7. Conclusion

We demonstrated that the transfer equation in one-dimensional spherical geometry with central symmetry can be reduced to a simple form. The derivative with respect to the direction cosine can be dropped out from the equation in conservation form without impeding the accuracy of the total intensity and the higher moments. We demonstrated this analytically and numerically by comparing the results with those obtained from the solution of the transfer equation with angular derivative in discrete ordinates formalism.

The gain from the proposed simplification of the transfer equation is the removal of the numerical error associated with the approximation of the angular derivative in this class of problems. Larger order of discrete ordinates can be used without running into significant round-off errors associated with the solution with angular derivative, and the CPU time is still much smaller than the time needed to work with the discrete ordinates method.

A set of coupled first order equations in $\phi$ and $\phi^*$ can be obtained by adding and subtracting (3.1) and (3.2) and making appropriate eliminations. The odd moments can then be computed from the summation of $w_i \mu_i^{2n-1} \phi_i^*$ and the even moments from the summation of $w_i \mu_i^{2(n-1)} \phi_i$. There is no apparent benefit from working with these first order equations at this point. There are as many $\phi$ and $\phi^*$ as there are $\psi^+$ and $\psi^-$ at every $r$. Whether a second order equation in $\phi$ would be more efficient than the first order equations in numerical computations that's yet to be seen.

The solution of the reduced transfer equation with scattering is on our slate for future studies.

## Acknowledgment.

This work was funded by the Department of Navy under a grant N00173-001-G010, and was solicited under a protocol with the Department of Physics and Astronomy at Louisiana State University.